\documentclass[journal]{IEEEtai}

\usepackage[colorlinks,urlcolor=blue,linkcolor=blue,citecolor=blue]{hyperref}

\usepackage{color,array}
\usepackage{booktabs}
\usepackage{float}
\usepackage{subfigure}  %插入多图时用子图显示的宏包
\usepackage{footnote}
\usepackage{cite}
\usepackage{amsmath,amsthm,amssymb,amsfonts}
\usepackage{algorithm}
\usepackage{algorithmicx}
\usepackage{algpseudocode}
\usepackage{graphicx}
\usepackage{textcomp}
\usepackage{xcolor}
\usepackage[T1]{fontenc}
\usepackage{bm}
\begin{document}

\title{MA2RL: Masked Autoencoders for Generalizable Multi-Agent Reinforcement Learning}

% \author{First A. Author, \IEEEmembership{Fellow, IEEE}, Second B. Author, and Third C. Author, Jr., \IEEEmembership{Member, IEEE}
% \thanks{This paragraph of the first footnote will contain the date on which you submitted your paper for review. It will also contain support information, including sponsor and financial support acknowledgment. For example, ``This work was supported in part by the U.S. Department of Commerce under Grant BS123456.'' }
% \thanks{The next few paragraphs should contain the authors' current affiliations, including current address and e-mail. For example, F. A. Author is with the National Institute of Standards and Technology, Boulder, CO 80305 USA (e-mail: author@boulder.nist.gov).}
% \thanks{S. B. Author, Jr., was with Rice University, Houston, TX 77005 USA. He is now with the Department of Physics, Colorado State University, Fort Collins, CO 80523 USA (e-mail: author@lamar.colostate.edu).}
% \thanks{T. C. Author is with the Electrical Engineering Department, University of Colorado, Boulder, CO 80309 USA, on leave from the National Research Institute for Metals, Tsukuba, Japan (e-mail: author@nrim.go.jp).}
% \thanks{This paragraph will include the Associate Editor who handled your paper.}}
\author{Jinyuan Feng,
        Min Chen,
        Zhiqiang Pu,
        Yifan Xu,
        Yanyan Liang
}

\markboth{Journal of IEEE Transactions on Artificial Intelligence, Vol. 00, No. 0, Month 2020}
{First A. Author \MakeLowercase{\textit{et al.}}: Bare Demo of IEEEtai.cls for IEEE Journals of IEEE Transactions on Artificial Intelligence}

\maketitle

\begin{abstract}

To develop generalizable models in multi-agent reinforcement learning, recent approaches have been devoted to discovering task-independent skills for each agent, which generalize across tasks and facilitate agents' cooperation. However, particularly in partially observed settings, such approaches struggle with sample efficiency and generalization capabilities due to two primary challenges: (a) How to incorporate global states into coordinating the skills of different agents? (b) How to learn generalizable and consistent skill semantics when each agent only receives partial observations? To address these challenges, we propose a framework called \textbf{M}asked \textbf{A}utoencoders for \textbf{M}ulti-\textbf{A}gent \textbf{R}einforcement \textbf{L}earning (MA2RL), which encourages agents to infer unobserved entities by reconstructing entity-states from the entity perspective. The entity perspective helps MA2RL generalize to diverse tasks with varying agent numbers and action spaces. Specifically, we treat local entity-observations as masked contexts of the global entity-states, and MA2RL can infer the latent representation of dynamically masked entities, facilitating the assignment of task-independent skills and the learning of skill semantics. Extensive experiments demonstrate that MA2RL achieves significant improvements relative to state-of-the-art approaches, demonstrating extraordinary performance, remarkable zero-shot generalization capabilities and advantageous transferability.

 % Additional rewards transform the original MTRL problem into a multi-objective MTRL problem, and the coupling relationship between the outputs of SP and ACP further complicates the optimization process. To solve this challenge, TSAC assigns a virtual expected budget to convert the multi-objective MTRL into a constrained single-objective formulation and then employs the Lagrangian method to transform a constrained single-objective optimization into an unconstrained one. The multiplier in the Lagrangian method automatically adjusts the weights during the training process, promoting cooperation between SP and ACP.
\end{abstract}
\begin{IEEEImpStatement}
The Current policies trained by Multi-Agent Reinforcement Learning (MARL) predominantly rely on meticulously designed structured environments, which considerably constrain the agents' generalization capabilities across multitasking and cross-task skill reuse. In this paper, we design a novel masked autoencoders for MARL to coordinate the skills of different agents and learn generalizable and consistent skill semantics when each agent only receives partial observations. Experimental results demonstrate that our proposed MA2RL framework significantly enhances both the asymptotic performance and generalization capabilities of the generalizable models. Specifically, MA2RL introduces masked autoencoders tailored for MARL, aimed at enhancing generalizable models. The framework holds promise for inspiring further explorations into the generalization of multi-agent reinforcement learning.
\end{IEEEImpStatement}

\begin{IEEEkeywords}
Multi-Agent reinforcement learning, generalization, self-supervised learning.
\end{IEEEkeywords}

\IEEEpeerreviewmaketitle
\section{Introduction}

% MPOWERING generalist robots through reinforcement learning is one of the essential targets of robotic learning.
\IEEEPARstart{I}{N} recent years, cooperative multi-agent reinforcement learning (MARL) has made significant progress in solving complex real-world tasks, such as autonomous driving, multi-robot systems, and sensor networks~\cite{xiao2023stochastic,wu2022deep,feng2024efficient,yun2022cooperative,okine2024multi,galvan2021neuroevolution}. Despite the success of MARL, most of the improvements are limited to single tasks. To improve the generalization of models across tasks, previous studies have primarily focused on solving varying agent numbers across tasks~\cite{hu2021updet,wang2022multi,xu2023improving,liu2023masked} and maintaining consistent semantics across tasks through the introduction of concepts like Skills/Roles/Subtasks~\cite{Decompose_Tian,zhang2022discovering}. 
While these approaches exhibit promising performance, they struggle with sample efficiency and generalization capabilities, particularly in partially observed settings.
% 部分可观带来的问题，引入自建督的表征学习方法,现有自建督方法的问题

Learning presentations that contain global state information can be an effective way to alleviate the aforementioned limitations. To facilitate the acquisition of informative representations, self-supervised learning (SSL) is commonly integrated with reinforcement learning (RL), especially in vision-based RL~\cite{laskin2020curl,zhu2022masked,yu2022mask,liu2022masked,kang2023sample,zhao2023learning,nguyen2021csnas}. As a critical technique in SSL, masked autoencoders (MAE) have made tremendous successes in computer vision (CV) and natural language processing (NLP)~\cite{devlin2018bert,he2022masked,brown2020language} for their ability to reconstruct masked tokens. Thus, recent works have explored the application of MAE in MAS \textit{e.g.,} MaskMA~\cite{liu2023masked}, MA2CL~\cite{song2023ma2cl} and MAJOR~\cite{feng2022joint}. While promising, they neglect entity-level information, limiting their applicability to a generalizable model structure. A detailed introduction to these methods is provided in Appendix \ref{Appendix:Related work} in the form of a related work section.
% masked autoencoder在不同领域应用的差别
% \yf{we explore the generality of MARL algorithms through masked autoencoders.} \yf{Is MAE generalizable learner in the context of MARL?}

In this work, we extend masked autoencoders (MAE)~\cite{he2022masked} in MARL to improve sample efficiency and generalization capabilities. As illustrated in ASN~\cite{wang2019action}, an agent's observation can be conducted as a concatenation of numerous entity-observations. Therefore, if treating each entity-observation as a token, the idea of MAE can be naturally applied in a generalizable model in MARL. Following the success of MAE in CV and NLP, we ask: \textit{how to extend MAE in the generalizable model in MARL?} As shown in Fig.\ref{fig:illustration_MAE}, we attempt to answer this question from the following insights:
\begin{figure*}[t]
    \centering
    \includegraphics[width=0.9\textwidth]{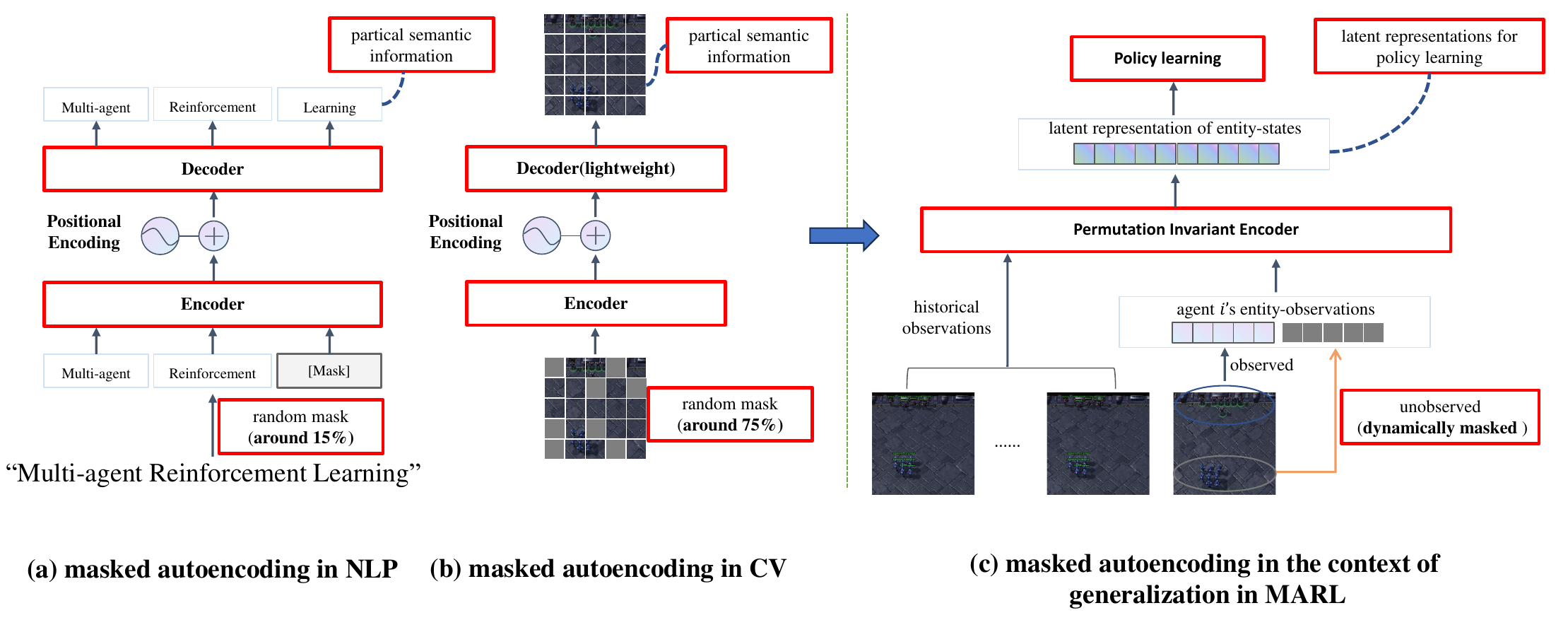}
    \caption{Illustration of MAE in different domains. (a) Masked autoencoding in NLP. (b) Masked autoencoding in CV. (c) Masked autoencoding in the context of generalization in MARL.
}\label{fig:illustration_MAE}
\end{figure*}
\begin{figure*}[]
    \centering
    \includegraphics[width=0.9\textwidth]{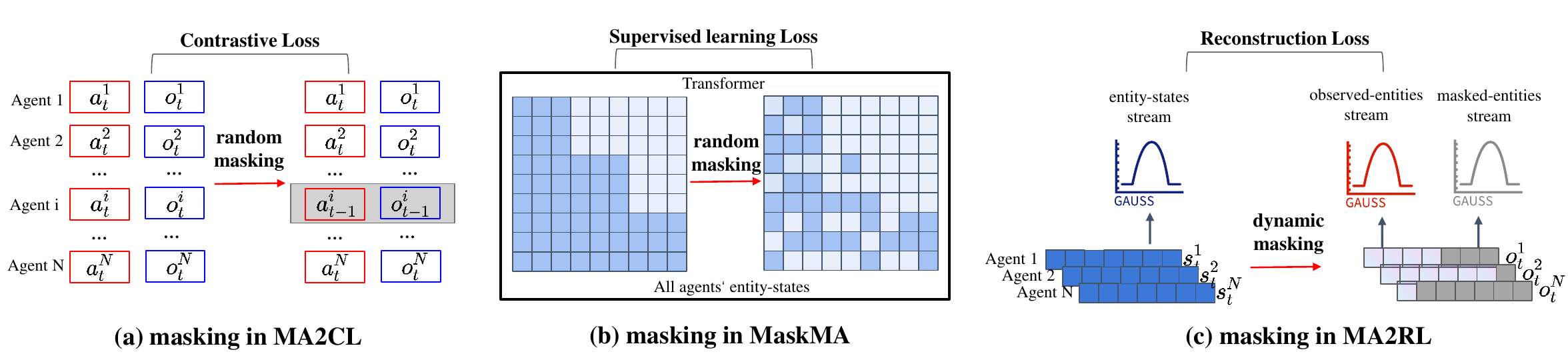}
    \caption{Simplified schematic diagram of typical methods for applying MAE in multi-agent systems (MA2CL, MaskMA and MA2RL). The figure shows a comparison of masking between MA2CL, MaskMA, and MA2RL. (a) random masking in MA2CL. (b) random masking in MaskMA. (c) dynamic masking in MA2RL}
\label{fig:simple_compare_mae}
\end{figure*}
\begin{enumerate}
  \item[(1)] \textbf{Inference at distinct scales}: In CV and NLP, a spatial contextual correlation exists between tokens. Therefore, it is necessary to incorporate position information to infer masked tokens or patches. Accordingly, MARL necessitates historical information for inference. In contrast to CV and NLP, MARL is commonly modeled as a partially observed Markov decision process (POMDP). Thus, masked entity-observations need to be inferred from distinct scales—"\textbf{temporal scale inference}". Additionally, the permutation invariance of entity-observations enables input embeddings to be independent of positional information.
  \item[(2)] \textbf{Dynamic masked ratio}: In CV and NLP, the masked ratio is fixed, which depends on information density. For example, BERT~\cite{devlin2018bert} predicts approximately 15\% of mask tokens, and MAE~\cite{he2022masked} masks a high portion of patches (around 75\%). Accordingly, in MARL, each agent's local entity-observation serves as a mask for the global entity-state. In contrast to CV and NLP, \textbf{dynamic masked ratio} in MARL calls for a generalizable model structure.
  \item[(3)] \textbf{Different reconstruction contents}: In CV and NLP, the optimization objective is to reconstruct the tokens which contain practical semantic information. Accordingly, MAE in MARL are designed to reconstruct the information of dynamically unobserved entities. In contrast to CV and NLP, MAE in MARL has \textbf{different reconstruction contents}, which aims at reconstructing the latent representation of dynamically unobserved entities and maximizing the cumulative return of policies. Policies rely solely on the latent representation of observations without requiring a decoder to explicitly reconstruct entity observations.
\end{enumerate}
% (1)嵌入的构成：在CV和NLP中，token和patch之间具有空间上的上下文相关性， 所以为了使模型能够利用序列的顺序，我们必须向序列中的表征注入一些关于相对或绝对位置的信息。而MARL通常被建模为马尔可夫决策过程，具有时间上的相关性，在对mask token的预测需要考虑历史轨迹的信息。此外，观测的置换不变性，使得嵌入表征无需位置信息。where temporal correlation needs to be considered for infering masked entity-observations
% (2)masked ratio:以bert和mae为例，mask的比例是固定的，具体的值取决于语义和信息的密度，例如，bert对单词mask的比例大概是15%，MAE对图片patch mask了75%。而在MARL中，智能体局部观测的entity-observation可以看作全局entity-state的mask，其中mask的ratio是随着与环境的交互动态变化的，取决于智能体观测的信息。
% (3)在CV和NLP中，通过decoder重建的pixel或者tokens有实际的语义信息，所以decoder在学习的潜在表示的语义级别方面起着关键作用。而在MARL中，策略学习一种观测和动作间的映射函数，所以无需使用decoder来重建实体观测信息。
% 我们的论文贡献。

Driven by this analysis, we propose a novel framework called \textbf{M}asked \textbf{A}utoencoders for \textbf{M}ulti-\textbf{A}gent \textbf{R}einforcement \textbf{L}earning (MA2RL), which implicitly reconstructs latent representation of dynamically masked entities for more comprehensive task-independent skills assignment and learning skill semantics in a global perspective. In MA2RL, entity-observations and entity-states are first mapped into a latent space using variational autoencoders (VAE). Subsequently, the MAE infers masked entities by reconstructing global entity-states. These reconstructed latent representations are then leveraged to select an appropriate task-independent skill. Finally, by reusing the decoder in VAE, the attentive action decoder learns skill semantics based on the latent representation of the masked entity-observations, the observed entity-observations, and the task-independent skill. Experimental results demonstrate the effective utilization of masked encoding in MA2RL, leading to improved asymptotic performance and generalization across a wide range of tasks. To the best of our knowledge, our work is the first attempt to employ MAE in the context of generalization in MARL. The contributions of our work are summarized as follows:
\begin{enumerate}
  \item[1)] We propose a novel framework called MA2RL, the first attempt to extend MAE in the context of generalization in MARL, which implicitly reconstructs the latent representation of dynamically masked entities for more comprehensive task-independent skills assignment and learning skill semantics in a global perspective.
  \item[2)] We explore MAE from the entity perspective, positioning it as orthogonal research to previous SSL methods. Surprisingly, we find that MAE are generalizable for MARL due to its remarkable ability to infer unobserved entities and learn high-capacity models.
  \item[3)] Through extensive experiments, we demonstrate that MA2RL achieves SOTA performance in various experimental settings, exhibiting remarkable generalization and transfer capabilities.
\end{enumerate}

To highlight the novelty of our work, we add a comparison between MA2RL and recent works that explore the application of MAE in MARL, such as MA2CL~\cite{song2023ma2cl} and MaskMA~\cite{liu2023masked}. As illustrated in Fig.~\ref{fig:simple_compare_mae}, MA2CL randomly masks the observations and actions of some agents at timestep $t$, and replaces them with the observations and actions from timestep $t-1$. This design idea aims to encourage learning representations to be both temporal and agent-level predictive. MA2CL mainly focused on the model’s asymptotic performance , neglecting its generalization across tasks. MaskMA is dedicated to building a single generalist agent with strong zero-shot capability. However, it merely applies MAE to the attention layer of the transformer, without considering the natural masking relationship between entity states and entity observations in MARL. Finally, MA2RL treats individual entity-observations as masked contexts of entity-states, naturally integrating MAE with a generalizable model in MARL. In this case, masking ratio is dynamically changing and determined by the environment, which hinders the direct application of MAE. To tackle this challenge, we develop a novel MAE in MA2RL. The main novelty of MAE in MA2RL is that, through utilizing Gaussian distribution and dividing agents' observation into two streams (observed-entities stream and masked-entities stream), MAE in MA2RL is endowed with generalization capability to infer the dynamically changing masked entities.

% We treat local entity-observations as masked context of global entity-states. And we extend MAE to MARL for reconstructing latent representation of dynamically masked entities, which enables agents to infer the global entity-states.
\section{Related work}
\label{Related work}
\subsection{Generalization in MARL}
There are two significant obstacles that limit the transferability and generalization capability in MARL: (a)varying state/observation/action spaces across tasks and (b)overfitting task-specific information. In response to the obstacle (a), ASN~\cite{wang2019action} decomposes an agent's observation into a composite of $n$ entity-observations. Subsequently, by aligning the entity-observations with entity-based actions, a model structure that is generalizable across tasks can be formulated. After that, UPDeT~\cite{hu2021updet} combines ASN with transformer blocks to improve the model's generalization. To convert obstacle (b), some previous works focused on knowledge transfer by learning presentations that capture the task-specific information~\cite{xu2023improving,qin2022multi,schafer2022learning,liu2019value}. Additionally, to accomplish interpretable cross-task decision-making, some works~\cite{zhang2022discovering,Decompose_Tian} turn to the concepts like skills/options/roles/subtasks in MARL for assistance. For example, DT2GS~\cite{Decompose_Tian} utilizes a scalable subtask encoder and an adaptive subtask semantic module to maintain consistent and scalable semantics across tasks. ODIS~\cite{zhang2022discovering} discovers task-invariant skills from multi-task offline data and improves generalization by coordinating discovered skills in unseen tasks. Overall, existing skill-based methods ~\cite{yang2022ldsa,chen2022multi,yang2024hierarchical,yang2019hierarchical,wang2020roma,wang2020rode,liu2022heterogeneous} in MARL suffer from deficiencies caused by partial observation, such as neglect of team awareness, or relaxation of the centralized training with decentralized execution (CTDE) constraint, which also impede the generalization and asymptotic performance in MARL. 
\subsection{Self-Supervised Learning in RL \& MARL}
Self-Supervised Learning (SSL) has made tremendous success in CV and NLP. Consequently, the idea of SSL is natural and applicable in RL to accomplish effective representation learning, particularly in vision-based RL environments. Substantial works~\cite{laskin2020curl,zhu2022masked,yu2022mask,yu2021playvirtual,yarats2021reinforcement,liu2024enhancing} construct auxiliary SSL objectives by considering the correlations among vision states or predicting the reward model and dynamic model in MDP. As a simple and effective technique, the paradigm of pretraining is also investigated to assist RL, such as generalizing to multiple downstream tasks~\cite{liu2022masked,schwarzer2021pretraining}, promoting exploration~\cite{liu2021behavior} and discovering skills\cite{liu2021aps}.To the best of our knowledge, the primary efforts on this direction have been paid on a single-agent setting, making MARL lags thus far. In ~\cite{shang2021agent,feng2022joint}, they focus on representation learning by predicting the future properties of all agents at the team level, such as location and observation. Besides, MA2CL~\cite{song2023ma2cl} encourages agents to take full advantage of temporal and agent-level information. And ACORM~\cite{hu2023attention} derive a contrastive learning objective to promote role representation learning. It appears that recent works in MARL focous solely on contrastive self-supervised learning, neglecting the potential of generative self-supervised learning.

\section{Preliminaries}
\label{Preliminaries}
\subsection{Multi-Agent Reinforcement Learning}
A MARL problem can be formulated as a decentralized partially observed Markov decision process (Dec-POMDP)~\cite{oliehoek2016concise}, which is described as a tuple $\langle n,\boldsymbol{S},\boldsymbol{A},P,R,\boldsymbol{O},\boldsymbol{\Omega},\gamma\rangle $, where $n$ represents the number of agents, $\boldsymbol{S}$ is the global state space. $\boldsymbol{A}$ is the action space. $\boldsymbol{O}=\{O_{i}\}_{i=1,\cdots,n}$ is the observation space. At timestep $t$, each agent $i$ receives an observation $o_{i}^t\in O_{i}$ according to the observation function $\boldsymbol{\Omega}(s^t,i):\boldsymbol{S}\to O_i$ and then selects an action $a_i^t\in\boldsymbol{A}$. The joint action $\boldsymbol{a}^t=(a_1^t,\ldots,a_n^t)$ is then applied to the environment, resulting in a transition to the next state $s^{t+1}$ and a global reward signal $r^{t}$ according to the transition function $P(s^{t+1}\mid s^{t},\boldsymbol{a}^t)$ and the reward function $R(s^t,\boldsymbol{a}^t)$. $\gamma\in[0,1]$ is the discount factor. The objective is to learn a joint policy $\pi$ that maximizes the expected cumulative reward $\mathbb{E}\left[\sum_{t=0}^{\infty}\gamma^{t}r^{t}\right|\pi]$.

\subsection{Centralized Training With Decentralized Execution}
Centralized Training with Decentralized Execution (CTDE) is a commonly employed architecture in MARL~\cite{lowe2017multi,rashid2020monotonic}. In CTDE, each agent utilizes an actor network to make decisions based on local observations. Additionally, the training process incorporates global information to train a centralized value function. The centralized value function provides a centralized gradient to update the actor network based on its outputs.

\subsection{Generalizable Model Structure in MARL}
To handle varying state/observation/action spaces, previous works like UPDeT~\cite{hu2021updet} and ASN~\cite{wang2019action} propose a generalizable model that treats all agents as entities. In such models, observation $o_i$ can be conducted as entity-observations: $[o_{i,1},o_{i,2},...,o_{i,m}]$, where $m$ denotes the number of all entities in the environment. Based on the criterion of whether entites can be observed, entity-observations can be splited into two subsets: observed entity-observations $o_{\mathrm{obs},i}$ and unobserved entity-observations $o_{\mathrm{mask},i}$. We denote the number of observed entities and masked entities as $n_{\mathrm{obs}}$ and $n_{\mathrm{mask}}$, respectively, and it holds that $m=n_{\mathrm{obs}}+n_{\mathrm{mask}}$.
Additionally, action space $\boldsymbol{A}$ can be decomposed into two subsets:$\boldsymbol{A}^{\mathrm{self}}$ containing actions that affect the environment or itself  and $\boldsymbol{A}^{\mathrm{out}}$ representing actions that directly interact with other entities.
\section{Method}
\label{sec:framework}
\begin{figure*}[t]
    \centering
    \includegraphics[width=\textwidth]{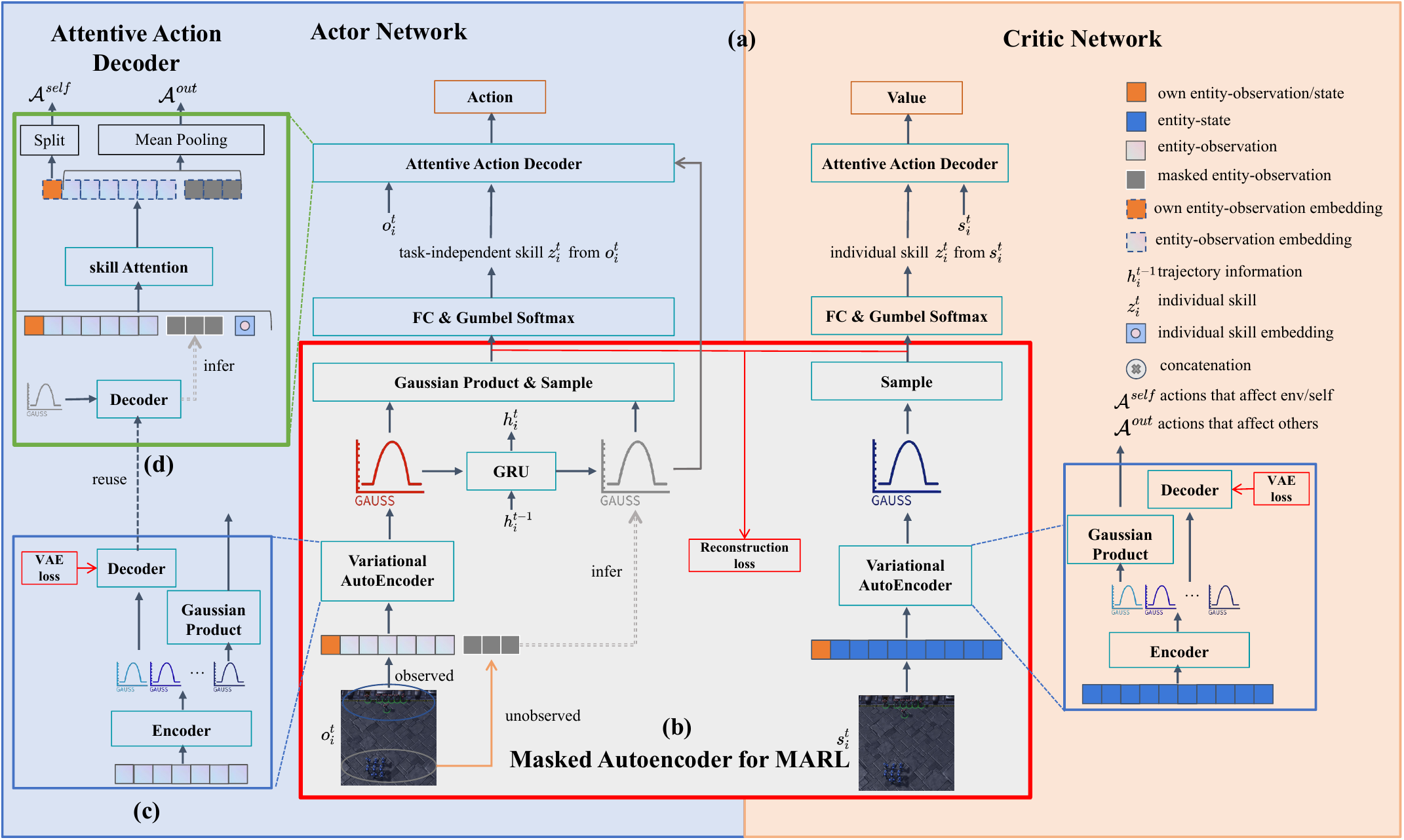}
    \caption{The network structure of MA2RL. (a) The overall architecture. (b) The stucture of variational autoencoder (VAE). (c) The details of masked autoencoder for MARL, where entity-observations can be regarded as a mask of the entity-states. (d) The attentive Action decoder that reuses the decoder in VAE to infer masked entity-observations for better action execution.
}
\label{fig:framework}
\end{figure*}

In this section, we will introduce our proposed framework MA2RL. Fig.\ref{fig:framework} illustrates the MA2RL framework, which consists of two components: masked autoencoders for MARL and attentive action decoder. The masked autoencoders for MARL enable agents to infer the global entity-states. And the attentive action decoder learns skill semantics from a global perspective. We outline the respective roles played by the two components in MA2RL:
\begin{itemize}
    \item masked autoencoders for MARL: it is utilized to incorporate global states into coordinating the skills of different agents, which plays a key role in asymptotic performance.
    \item action decoder: it is utilized to encompass the impact of inferred unobserved entities for achieving generalizable and comprehensive skill semantics, which play a key role in generalization.
\end{itemize}

Each component is introduced in detail in the following subsections.

\subsection{Masked Autoencoders for MARL}\label{subsec:MAE}
For agent $i$, its global entity-states at timestep $t$ can be conducted as : $s_i^t = [s_{i,1}^t,s_{i,2}^t,...,s_{i,m}^t]$, where $m$ denotes the total number of entities in the environment. Due to the partial observability of MAS, agent $i$ at timestep $t$ can only access local entity-observations $o_i^t = [o_{i,1}^t,o_{i,2}^t,...,o_{i,m}^t]$, where unobserved entities are zero padding. The connection between the global entity-states $s_i^t$ and the local entity-observation $o_i^t$ is established through a mask function $\bm{M}_i^t$, expressed as follows:
\begin{equation}
    \begin{split}
        {\bm{M}_i^t}&=\{M_{i1}^t,M_{i2}^t,\ldots,M_{ij}^t,\ldots,M_{im}^t\},M_{ij}^t=1 ~\mathrm{or}~ 0.\\
        o_{i}^t&={\bm{M}_i^t}\cdot s_i^t
    \end{split}
\end{equation}
If $M_{ij}^t=0$, the $j$-th entity at timestep $t$ is unobserved for agent $i$. We reconstruct the entity-states $s_i^t$ by leveraging information from the historical trajectory and currently observed entity-observations.

As illustrated in Fig.\ref{fig:framework}(b), we employ two variational autoencoders (VAEs) to derive latent representations from the local entity-observations and the global entity-states. In the VAEs, the encoder transforms entity-observations and entity-states into a Gaussian distribution in the latent space, while the decoder reconstructs entity-observations and entity-states by decoding samples from the Gaussian distributions. There are two key advantages to selecting a normal distribution as the encoded distribution: 1) Utilizing Gaussian Product with Gaussian distribution enables us to obtain fixed-dimensional latent representations of entity-observations. 2) The Gaussian Product naturally satisfies the permutation invariance property of entity-observations. Specifically, we employ two VAEs, which encoders and decoders parameterized by $(\theta_{\mathrm{oe}} , \theta_{\mathrm{od}}),(\theta_{\mathrm{se}},\theta_{\mathrm{sd}})$ to encode the entity-observations and entity-states into Gaussian distributions respectively. The VAE models are trained using the $loss_{\mathrm{VAE}}$ objective function:
\begin{figure*}[t]
    \centering
    \includegraphics[width=\textwidth]{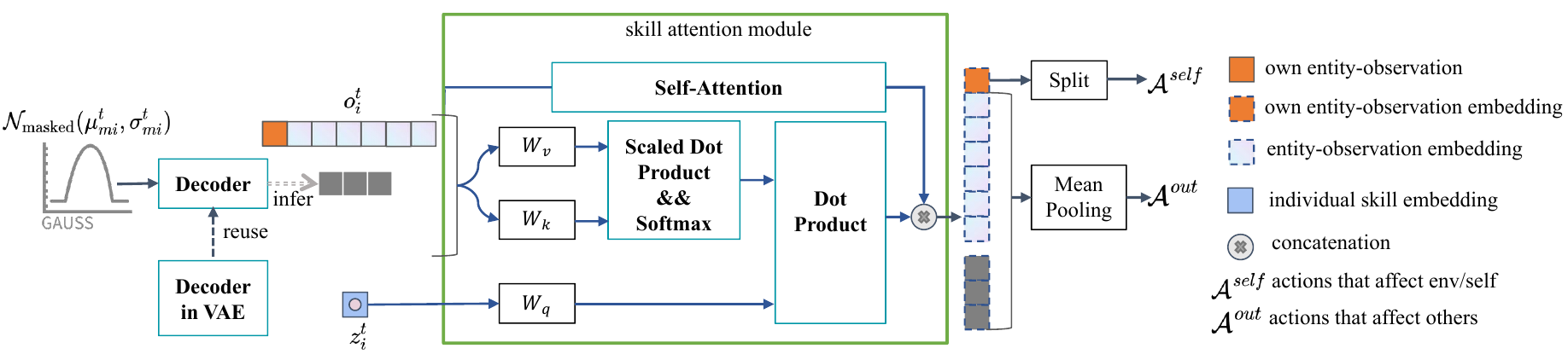}
    \caption{attentive action decoder. The attentive action decoder utilizes the latent representations of all masked entity-observations to infer the information of masked entities and then applies a skill attention module to obtain actions.
}
\label{fig:attention_decoder}
\end{figure*}
\begin{equation}
\begin{split}
(\mu_{\mathrm{o},ij}^t,\sigma_{\mathrm{o},ij}^t)&=f_{\theta_{\mathrm{oe}}}(o_{ij}^t),j=1,2,...,m, o_{ij}^t \in o_{\mathrm{obs},i}^t\\
\mathcal{L}_{\mathrm{VAE_o}}&=\|o_{ij}^t-f_{\theta_{\mathrm{od}}}(x \sim \mathcal{N}(\mu_{\mathrm{o},ij}^t, \sigma_{\mathrm{o},ij}^t))\|^2\\
(\mu_{\mathrm{s},ij}^t,\sigma_{\mathrm{s},ij}^t)&=f_{\theta_{\mathrm{se}}}(o_{ij}^t),j=1,...,m\\
\mathcal{L}_{\mathrm{VAE_s}}&=\|s_{ij}^t-f_{\theta_{\mathrm{sd}}}(x \sim \mathcal{N}(\mu_{\mathrm{s},ij}^t, \sigma_{\mathrm{s},ij}^t))\|^2
\label{eq:(2)}
\end{split}
\end{equation}
$(f_{\theta_{\mathrm{od}}},f_{\theta_{\mathrm{od}}}),(f_{\theta_{\mathrm{se}}},f_{\theta_{\mathrm{sd}}})$ respectively denotes the encoders and decoder of two VAEs. Thus, the latent representation of all observed entity-observations and entity-states, respectively denoted as Gaussian distributions $\mathcal{N}_{\mathrm{observed}}(\mu_{\mathrm{o}i}^t, \sigma_{\mathrm{o}i}^t)$, $\mathcal{N}_{\mathrm{states}}(\mu_{\mathrm{s}i}^t,\sigma_{\mathrm{s}i}^t)$, are constructed using Gaussian Product:
\begin{equation}
\begin{split}
\mathcal{N}_{\mathrm{observed}}(\mu_{\mathrm{o}i}^t,\sigma_{\mathrm{o}i}^t)&=\prod_{j=1}^{n_{\mathrm{obs}}}\mathcal{N}(\mu_{\mathrm{o},ij}^t,\sigma_{\mathrm{o},ij}^t)\\
\mathcal{N}_{\mathrm{states}}(\mu_{\mathrm{s}i}^t,\sigma_{\mathrm{s}i}^t)&=\prod_{j=1}^m\mathcal{N}(\mu_{\mathrm{s},ij}^t,\sigma_{\mathrm{s},ij}^t)
\label{eq:(3)}
\end{split}
\end{equation}

\subsection{Attentive action decoder}
Subsequently, as depicted in Fig.\ref{fig:framework}(b), we employ a GRU, parameterized by $f_{\theta_h}$, to infer the latent representations of all masked entity-observations $\mathcal{N}_{\mathrm{masked}}(\mu_{mi}^t, \sigma_{mi}^t)$. The GRU takes the latent representation of all observed entity-observations $\mathcal{N}_{\mathrm{observed}}(\mu_{oi}^t, \sigma_{oi}^t)$ and the previous trajectory information $h_i^{t-1}$ as inputs. This process can be formulated as follows:
\begin{equation}
    \mathcal{N}_{\mathrm{masked}}(\mu_{mi}^t, \sigma_{mi}^t), h_i^t=f_{\theta_h}(x \sim \mathcal{N}_{\mathrm{observed}}(\mu_{oi}^t,\sigma_{oi}^t),h_i^{t-1})
    \label{eq:(4)}
\end{equation}
To merge the encoded and inferred results, we employ Gaussian Product to merge the two Gaussian distributions and obtain an integrated representation of the entity-observations:
\begin{equation}
        \mathcal{N}_{\mathrm{integration}}(\mu_{i}^t,\sigma_{i}^t)=\mathcal{N}_{\mathrm{observed}}(\mu_{oi}^t,\sigma_{oi}^t)\cdot\mathcal{N}_{\mathrm{masked}}(\mu_{mi}^t, \sigma_{mi}^t)
    \label{eq:(5)}
\end{equation}
By utilizing the integrated representation of the entity-observations and the latent representation of entity-states, we construct a reconstruction loss function to optimize the parameters of the GRU, enabling it to infer the latent representations of the masked entity-observations. The reconstruction loss is formulated as follows:
\begin{equation}
\begin{split}
\mathcal{L}_{\mathrm{reconstruct}}=\|(y \sim \mathcal{N}_{\mathrm{states}}(\mu_{si}^t,\sigma_{si}^t))-\\
(x \sim \mathcal{N}_{\mathrm{integration}}(\mu_{i}^t,\sigma_{i}^t))\|^2
\end{split}
\label{eq:(6)}
\end{equation}
$x$ and $y$ are sampled from Gaussian distributions $\mathcal{N}_{\mathrm{states}}$ and $\mathcal{N}_{\mathrm{integration}}$, respectively. By minimizing the reconstruction loss $\mathcal{L}_{\mathrm{reconstruct}}$, we can train both the encoder and the GRU to effectively extract informative representations and recover the latent representation of masked entity-observations.

By introducing historical information, MAE in MARL can attempt to figure out the enemy's representation from the following two aspects: (1) Historical opponent information observed (e.g., if an agent observes the enemy's information at time steps 5 to 10, then after step 11, the opponent's representation can be predicted based on historical information). (2)Currently observed information about allies (e.g., predicting the opponent's representation through the historical trajectory or current behavior of teammates).

According to the masked autoencoders for MARL, we obtain the integrated representation of the entity-observations $\mathcal{N}_{\mathrm{integration}}(\mu_{i}^t,\sigma_{i}^t)$, which contains the observed entities' information and the inferred masked entities' information. Based on the integrated representation of the entity-observations, we utilize the Gumbel-Softmax trick to assign an individual skill $z_i^t$:
\begin{equation}
z_i^t=\mathrm{Gambel-Softmax}(x \sim \mathcal{N}_{\mathrm{integration}}(\mu_{i}^t,\sigma_{i}^t))
\label{eq:skill}
\end{equation}
Inspired by DT2GS~\cite{Decompose_Tian} and ACORM~\cite{hu2023attention}, which utilize the attention mechanism~\cite{vaswani2017attention} to incorporate subtasks/roles into the action decoder, we introduce the concept of task-independent skills to facilitate the training process of the attentive action decoder. In DT2GS~\cite{Decompose_Tian}, subtask semantics refer to the effects of an agent on observed entities. While in MA2RL, task-independent skill semantics not only refer to the impact of observed entities but also encompass the impact of inferred unobserved entities for achieving generalizable and comprehensive skill semantics.

\begin{algorithm}[]
\caption{MA2RL}
\begin{algorithmic}[1]
\Statex \textbf{Input:} The VAE parameters $(\theta_{\mathrm{oe}},\theta_{\mathrm{od}})$ for the masked autoencoder of actor $\pi$; The VAE parameters $(\theta_{\mathrm{se}},\theta_{\mathrm{sd}})$ for the masked autoencoder of critic $V$; the parameters $\theta_{\mathrm{mae}}$ for mask encoder in $\pi$; the parameters $\theta_{\mathrm{de}}$ for attentive action decoder in $\pi$; the parameters $\theta^V$ for critic $V$.
\State{Initialize $(\theta_{\mathrm{oe}},\theta_{\mathrm{od}})$,$(\theta_{\mathrm{se}},\theta_{\mathrm{sd}})$, $V$} % Discriminator parameters 
\State{Initialize the total timesteps $T$ of an episode; the replay buffer $D$}
\State{Initialize $step=0$}
\For{$episode =0,1,\dots,K$}
\State{initialize actor RNN states $h_{1,\pi}^0,\dots,h_{n,\pi}^0$}
\State{initialize critic RNN states $h_{1,V}^0,\dots,h_{n,V}^0$}
\State{$\tau=\left[\right]$}
    \For{$timestep=1$ to $T$}
        \For{$i=0$ to $n$, agent $i$}
        \State{$\mathcal{N}_{\mathrm{masked}}(\mu_{i}^t,\sigma_{i}^t)$,$\mathcal{L}_{\mathrm{reconstruct}},h_i^{t}=\pi_{\mathrm{mae}}(o_i^t,h_i^t;\theta_{\mathrm{mae}})$,$ \triangleright$Call for Algorithm \ref{mae4marl}}
        \State{Choose an individual skill $z_i^t$ by by Formula(\ref{eq:skill})}
        \State{$a_i^t=\pi_{de}(o_i^t,\mathcal{N}_{\mathrm{masked}}(\mu_{mi}^t, \sigma_{mi}^t),z_i^t;\theta_{\mathrm{de}})$,$\triangleright$Call for Algorithm \ref{action_decoder}}
        \EndFor 
        \State{Execute actions $a^t$, then acquire $r^t,s^{t+1},o^{t+1}$}
        \State{$\tau+=[s^t,\boldsymbol{o^t},\boldsymbol{h_\pi^t},\boldsymbol{h_V^t},\boldsymbol{k^t},\boldsymbol{a^t},r^t,s^{t+1},\boldsymbol{o^{t+1}}]$}
    \EndFor
    \State{$D=D\cup\left(\tau\right)$}
    \For{mini-batch $k=1,...K$}
    \State{Sample data from $D$ to update $\pi$ by minimizing actor loss}
    \State{Sample data from $D$ to update $V$ by minimizing critic loss}
    \State{Sample data from $D$ to update $(\theta_{\mathrm{oe}},\theta_{\mathrm{od}});(\theta_{\mathrm{se}},\theta_{\mathrm{sd}})$ by minimizing $\mathcal{L}_{\mathrm{reconstruct}}$ }
    \EndFor
\EndFor
\end{algorithmic}
\label{alg}
\end{algorithm}

\begin{algorithm}[]
\caption{Masked autoencoders for MARL}
    \begin{algorithmic}[1]
\Statex \textbf{Input:} The VAE parameters $(\theta_{\mathrm{oe}},\theta_{\mathrm{od}})$ for masked autoencoder of actor $\pi$, The VAE parameters $(\theta_{\mathrm{se}},\theta_{\mathrm{sd}})$ for masked autoencoder of critic $V$, agent $i$'s observation $o_i^t$, agent $i$'s state $s_i^t$ and previous trajectory information $h_i^{t-1}$
\State{Encode the entity-observations and entity-states into $(\mu_{\mathrm{o},ij}^t,\sigma_{\mathrm{o},ij}^t);(\mu_{\mathrm{s},ij}^t,\sigma_{\mathrm{s},ij}^t),j=1,...,m$ by Formula(\ref{eq:(2)})}
\State{Compute $\mathcal{N}_{\mathrm{observed}}(\mu_{oi}^t,\sigma_{oi}^t);\mathcal{N}_{\mathrm{states}}(\mu_{si}^t,\sigma_{si}^t)$ by Formula(\ref{eq:(3)})}
\State{Infer the latent representation of all masked entity-observations $\mathcal{N}_{\mathrm{masked}}(\mu_{mi}^t, \sigma_{mi}^t)$ and $h_i^{t}$ by Formula(\ref{eq:(4)})}
\State{Obtain an integrated representation of the entity-observations $\mathcal{N}_{\mathrm{integration}}(\mu_{i}^t,\sigma_{i}^t)$ by Formula(\ref{eq:(5)})}
\State{Compute the reconstruction loss $\mathcal{L}_{\mathrm{reconstruct}}$ by Formula(\ref{eq:(6)})}
\State{\textbf{Return} $\mathcal{N}_{\mathrm{masked}}(\mu_{mi}^t, \sigma_{mi}^t)$,$\mathcal{L}_{\mathrm{reconstruct}},h_i^{t}$}
\end{algorithmic}
\label{mae4marl}
\end{algorithm}

\begin{algorithm}[]
\label{algorithm2}
\caption{Attentive action decoder}
    \begin{algorithmic}[1]
\Statex {\textbf{Input:} The VAE parameters $(\theta_{\mathrm{oe}},\theta_{\mathrm{od}})$ for masked autoencoder of actor $\pi$, agent $i$'s observation $o_i^t$ and skill $z_i^t$ at timestep $t$}
\State{Obtain the enhanced entity-observation $o_{\mathrm{enh},i}^t$ by Formula(\ref{eq:(7)})}
\State{Compute the embedding of the self-attention part $\tau_{i,\mathrm{self}}^t$ by Formula(\ref{eq:(8)})}
\State{Compute the embedding of the skill-based attention part $\tau_{i,skill}^t$ by Formula(\ref{eq:(9)})(\ref{eq:(10)})}
\State{Sample action $a_i^t$ from actions' probability distribution $P$ by Formula(\ref{eq:(11)}}
\State{\textbf{Return} action $a_i^t$}
\end{algorithmic}
\label{action_decoder}
\end{algorithm}
% 具体讲一下skill的概念 参考别的文章怎么讲的
As illustrated in Fig.\ref{fig:framework}(d), the attentive action decoder takes the local entity-observation $o_i = [o_{i1},o_{i2},...,o_{im}]$, the individual skill $z_i^t$ and the latent representations of all the masked entity-observations $\mathcal{N}_{\mathrm{masked}}(\mu_{mi}^t, \sigma_{mi}^t)$ as inputs and generates actions. We also provide a detailed description of the structure in Fig.\ref{fig:attention_decoder}. The latent representation of all masked entity-observations $\mathcal{N}_{\mathrm{masked}}(\mu_{mi}^t, \sigma_{mi}^t)$ are trained by MAE in Section\ref{subsec:MAE} and encompass speculative information about all unobserved entities. It is worth noting that the latent representation of all masked entity-observations is used to train a policy, thereby eliminating the need to explicitly reconstruct the observations of masked entities individually. Specifically, we utilize the inferred masked entity-observations by reusing the decoder parameterized by $f_{\theta_{\mathrm{od}}}$ in VAE:
\begin{equation}
\begin{split}
o_{\mathrm{enh},ij}^t&=\left\{\begin{array}{cc}
f_{\theta_{\mathrm{od}}}(m_{ij}^t \sim \mathcal{N}_{\mathrm{masked}}(\mu_{mi}^t, \sigma_{mi}^t)) & \text { if } o_{ij}^t \in o_{\mathrm{mask},i}^t \\
o_{ij}^t & \text { otherwise }
\end{array}\right.\\
o_{\mathrm{enh},i}^t&=[o_{\mathrm{enh},i1}^t,o_{\mathrm{enh},i2}^t,...,o_{\mathrm{enh},im}^t]
\end{split}
% o_{i1}, o_{i2},\ldots,o_{ij} \sim \mathcal{N}_{masked}(\mu_{mi}^t, \sigma_{mi}^t) ,j=1,2,...,n_{mask}, o_{ij} \in o_{mask,i}
\label{eq:(7)}
\end{equation}
$m_{ij}^t$ is a hidden embedding sampled from $\mathcal{N}_{\mathrm{masked}}(\mu_{mi}^t, \sigma_{mi}^t)$. After obtaining the enhanced entity-observation $o_{\mathrm{enh},i}^t$ by integrating the observed entity-observation and the inferred masked entity-observation, we introduce a skill attention module to learn skill semantics and enhance representational capacity and generalization. The skill attention module consists of two parts: self-attention to observed entities and skill-based attention. Specifically, in the self-attention part, we use the enhanced entity-observations as queries, keys and values:$Q_i^t=W_Qo_{\mathrm{enh},i}^t,K_i^t=W_Ko_{\mathrm{enh},i}^t,V_i^t=W_Vo_{\mathrm{enh},i}^t$, where $W_Q, W_K, W_V$ are parameter matrices for linear transformation. Formally, we calculate the embeddings of the self-attention part of agent $i$ at timestep $t$ as $\tau_{i,\mathrm{self}}^t$:
\begin{equation}
\tau_{i,\mathrm{self}}^t=\mathrm{softmax}(\frac{Q_i^tK_i^t}{\sqrt{d_k}})V_i^t
\label{eq:(8)}
\end{equation}
where $d_k$ represents the feature embedding of $K_i^t$ and $1/\sqrt{d_k}$ is a factor that scales the dot-product attention. Analogously, we compute the embedding of the skill-based attention part, denoted as $\tau_{i,\mathrm{skill}}^t$, by introducing the  individual skill $z_i^t$ as follows:
\begin{equation}
\tau_{i,\mathrm{skill}}^t=\sum_{j=1}^m\alpha_jv_{ij}^t=\sum_{j=1}^m\alpha_j\cdot W_Vo_{\mathrm{enh},ij}^t,
\label{eq:(9)}
\end{equation}
The attention weight $\alpha_j$ quantifies the relevance between the $j$-th entity-observation in $o_{\mathrm{enh},i}$ and the individual skill $z_i^t$. We apply a MLP and a softmax function to obtain the weight as:
\begin{equation}
    \alpha_j=\frac{\exp\left(\frac1{\sqrt{d_k}}\cdot W^Q\mathrm{MLP}(z_i^t) \cdot\left(W^Ko_{\mathrm{enh},ij}^t\right)^\top\right)}{\sum_{k=1}^m\exp\left(\frac1{\sqrt{d_k}}\cdot W^Q\mathrm{MLP}(z_i^t) \cdot\left(W^Ko_{\mathrm{enh},ik}^t\right)^\top\right)},
    \label{eq:(10)}
\end{equation}
In practice, we employ multi-head attention (MHA) to stabilize the learning process and collectively focus on information from different representation subspaces. Subsequently, we calculate the probability of each action/value by concatenating the embedding of the self-attention part $\tau_{i,\mathrm{self}}^t$ and the embedding of the skill-attention part $\tau_{i,\mathrm{skill}}^t$:
\begin{equation}
    P(a|o_i^t)= W[\tau_{i,\mathrm{self}}^t,\tau_{i,\mathrm{skill}}^t]
    \label{eq:(11)}
\end{equation}
$W$ is parameter matrices for linear transformation. The attentive action decoder enables us to flexibly leverage the latent representation of all masked entity-observations, thus alleviating performance degradation and improving generalization caused by partial observability. The skill attention module enhances the generalization of policies by introducing individual skills that can be generalized across tasks.

\subsection{Objective}
In this section, we present the training process of  MA2RL. The details of MA2RL is as shown in Algorithm 1, with the details of the two main modules, MAE and the action decoder, presented in Algorithms 2 and 3 respectively.

\begin{figure*}[ht]
    \centering
    \includegraphics[width=0.9\textwidth]{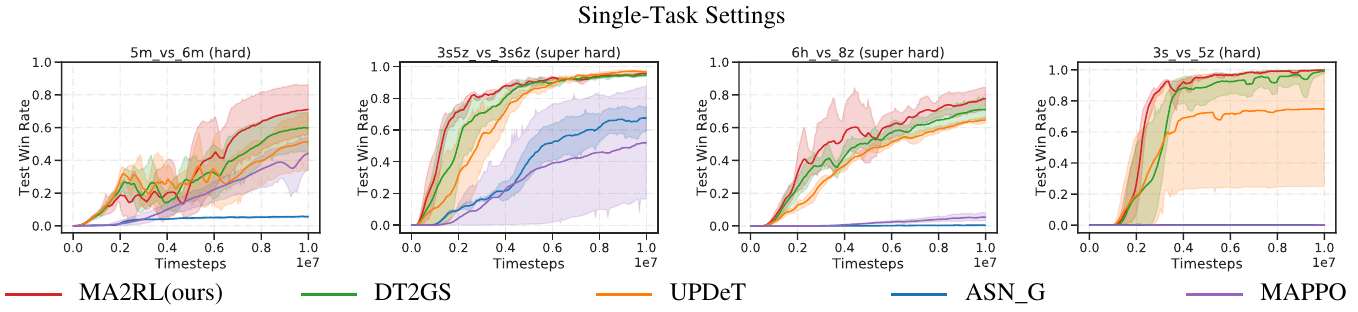}
    \caption{The performance of MA2RL and baselines, including DT2GS, UPDeT, ASN\_G, and MAPPO, are compared in the Single-Task settings. The evaluation is conducted on 2 hard tasks (5m\_vs\_6m, 3s\_vs\_5z) and 2 superhard tasks (3s5z\_vs\_3s6z, 6h\_vs\_8z).}
\label{fig:single_task}
\end{figure*}
\begin{figure}[h]
    \centering
    \includegraphics[width=0.5\textwidth]{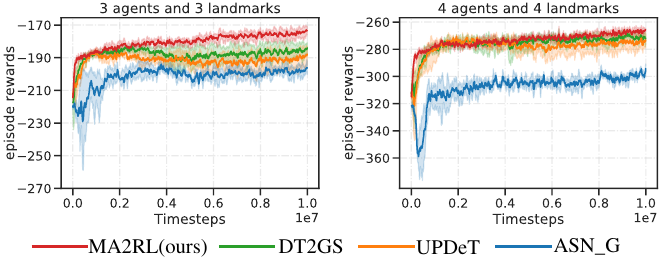}
    \caption{The asymptotic performance of MA2RL and baselines in MPE.}
\label{fig:asymptotic_mpe}
\end{figure}
\begin{figure*}[ht]
    \centering
    \includegraphics[width=0.9\textwidth]{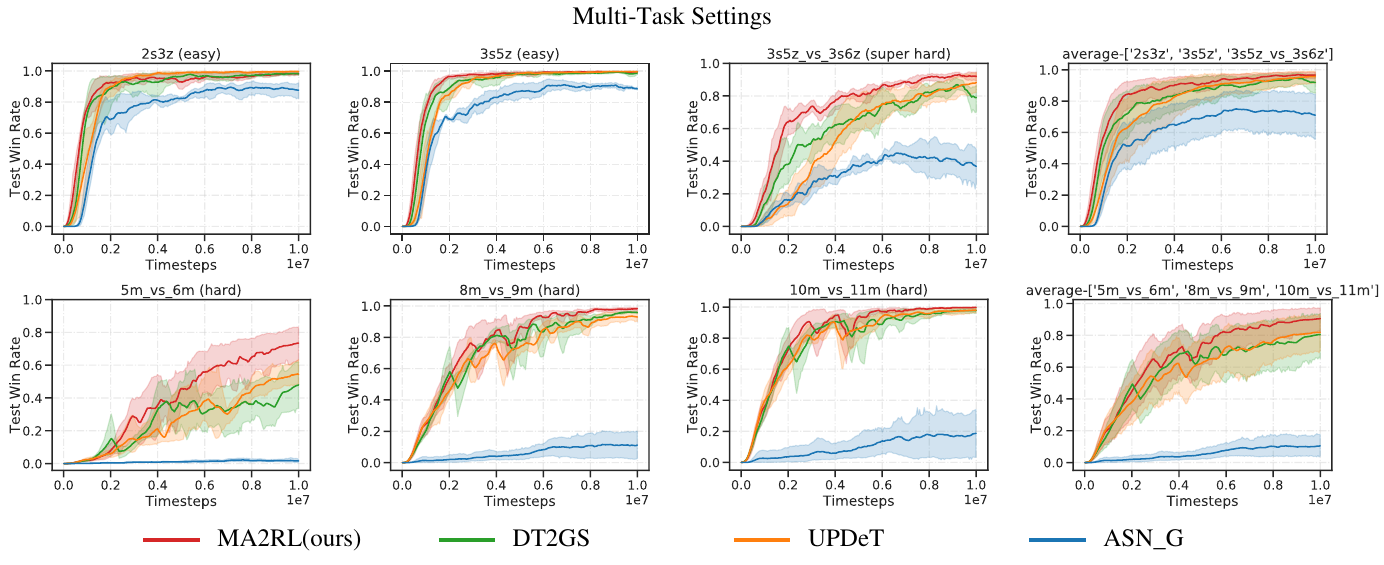}
    \caption{The performance of MA2RL and baselines, including DT2GS, UPDeT, ASN\_G, are compared in the Multi-Task settings. The evaluation is conducted on 2 multi-task problems with different distributions of difficulty: (2s3z, 3s5z, 3s5z\_vs\_3s6z),(5m\_vs\_6m, 8m\_vs\_9m, 10m\_vs\_11m) }.
\label{fig:multi_task}
\end{figure*}

\section{Experiments}
\label{sec:exp}
This section presents the experimental setup used to evaluate our proposed MA2RL and compare it against baselines. We conduct various multi-agent tasks to answer the following questions: (\textbf{Q1}) Can MA2RL effectively utilize global information and facilitate learning efficiency in complex multi-agent tasks and multi-task problems? (Section ~\ref{subsec:multi-task}) (\textbf{Q2}) Are masked autoencoders generalizable learners for MARL? And can the reduction of partial observability impact enhance the generalization and transferability of models?? (Section ~\ref{subsec:generalization})(\textbf{Q3}) How can we better leverage the latent representation of all masked entity-observations to enhance MA2RL's performance? (Section ~\ref{subsec:ablation})
\subsection{Experimental Setup}
\label{sec:exp_setup}
To evaluate the effectiveness of MA2RL, we conduct experiments with different scenarios and settings on two classical partially observed environments: StarCraft Multi-Agent Challenge (SMAC)~\cite{samvelyan2019starcraft} and MPE~\cite{lowe2017multi}. In SMAC and MPE, the policy network controls agents to perform decentralized actions. Thus, partial observability, to some extent, hinders the performance and generalization of the policy network. Our approach not only focuses on asymptotic performance in single-task settings but also on multi-task settings, as well as zero-shot generalization and transferability.

\textbf{Baselines:}We compare MA2RL with DT2GS~\cite{Decompose_Tian}, UPDeT~\cite{hu2021updet} and ASN\cite{wang2019action}. MAPPO~\cite{yu2022surprising} is added to the asymptotic performance experiment in single-task settings. DT2GS is chosen because it is the SOTA method and possesses sound zero-shot generalization capability across tasks by maintaining consistent yet scalable semantics. It is also used in the attentive action decoder in MA2RL. UPDeT develops a multi-agent framework based on the transformer block to adapt to tasks with varying observation/state/action spaces. ASN considers the semantic difference of actions and forms a foundation of generalizable models in MARL. Similar to the settings in DT2GS~\cite{Decompose_Tian}, our experiments use "ASN\_G" to denote the generalizable ASN. The criteria for choosing baselines depend on whether they can be applied across tasks in MARL or state-of-the-art methods.

\textbf{Architecture, Hyperparameters, and Infrastructure}:We implement MA2RL based on the codebase of MAPPO and DT2GS. The hyperparameters for MA2RL and baselines are presented in Table ~\ref{network_hyper} and Table ~\ref{algorithm_hyper}. With the model hyperparameters and training configurations above, a single job of MA2RL tasks up to 6 hours training on most of the SMAC maps, on a single machine of AMD EPYC 7742 CPU@2.25GHz with 64 CPU cores and eight RTX 3090 GPU. In practice, we use cluster with similar hardware to launch multiple jobs in parallel.
\begin{table*}[!ht]
    \centering
    \caption{Network hyperparameters used for MA2RL and baselines}
    \begin{tabular}{ccc}
    \hline
        \textbf{Hyperparameters} & \textbf{Value} & \textbf{Algorithms} \\ \hline
        hidden layer dimension of Encoder & 8 & MA2RL, DT2GS \\ 
        MLP's hidden layer dimension  & 64 & MA2RL, DT2GS, UPDeT, ASN\_G, MAPPO \\ 
        attention's hidden layer dimension & 64 & MA2RL, DT2GS, UPDeT, ASN\_G \\ 
        attention heads & 3 & MA2RL, DT2GS, UPDeT, ASN\_G \\ 
        number of subtasks/skills & 4 & MA2RL, DT2GS \\ \hline
    \end{tabular}
    \label{network_hyper}
\end{table*}

\begin{table*}[!ht]
    \centering
    \caption{Algorithm hyperparameters used for MA2RL and baselines}
    \begin{tabular}{cccccc}
    \hline
        \textbf{Hyperparameters} & \textbf{Value} & \textbf{Hyperparameters} & \textbf{Value} & \textbf{Hyperparameters} & \textbf{Value} \\ \hline
        optimizer & Adam & huber delta & 10 & use orthogonal & True \\ 
        learning rate & 0.0005 & clip param & 0.2 & clip param & 0.2 \\ 
        rollout threads & 24 & gae lambda & 0.95 & value\_loss\_coef & 1 \\ 
        num mini batch & 8 & use gae & True & gamma & 0.99 \\ 
        use valuenorm & True & gain & 0.01 & gae\_lambda & 0.95 \\ \hline
    \end{tabular}
    \label{algorithm_hyper}
\end{table*}

\subsection{Performance on single-task and multi-task settings}
\label{subsec:multi-task}
\begin{figure*}[ht]
    \centering
    \includegraphics[width=0.8\textwidth]{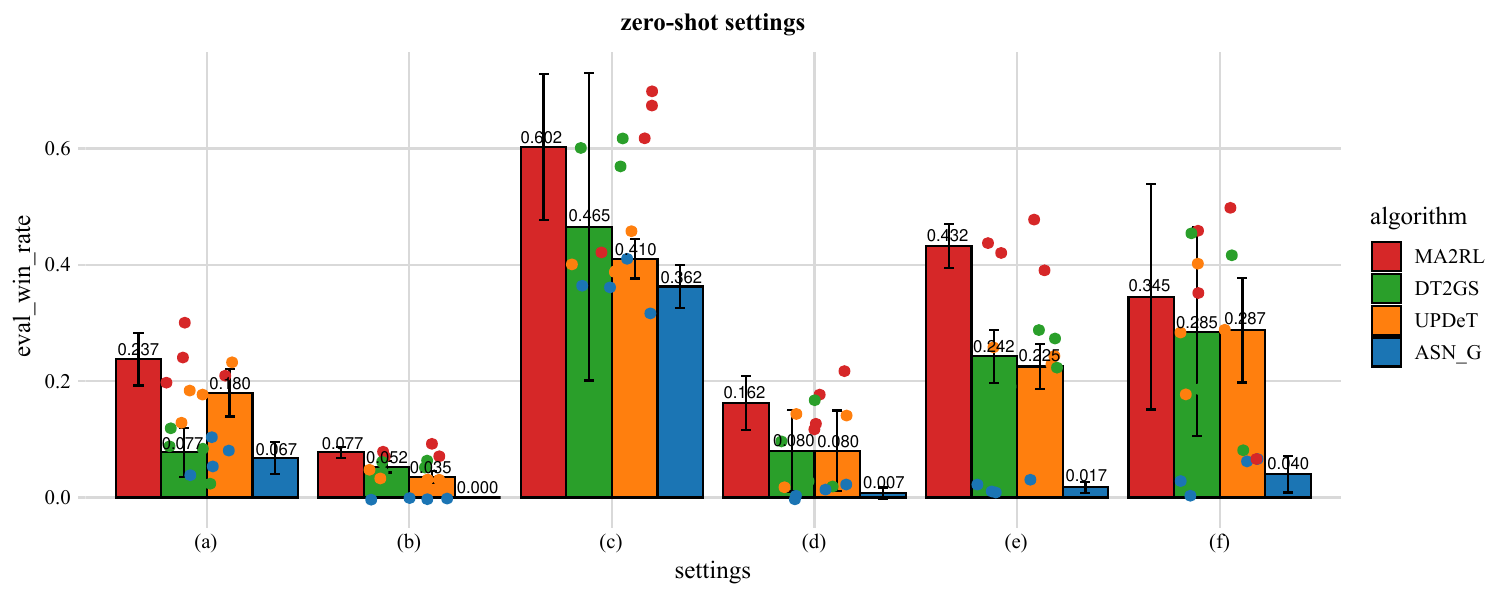}
    \caption{The zero-shot generalization capability of MA2RL and its baselines, including DT2GS, UPDeT, ASN\_G, was compared across various source and target tasks. The evaluation was conducted on 6 zero-shot settings and the horizontal axis represent the source task $\rightarrow$ the target task, where (a) 2s3z$\rightarrow$3s5z, (b) 3s\_vs\_4z$\rightarrow$3s\_vs\_5z, (c) 3s5z$\rightarrow$3s5z\_vs\_3s6z , (d) 8m\_vs\_9m$\rightarrow$5m\_vs\_6m, (e) 10m\_vs\_11m$\rightarrow$8m\_vs\_9m, (f) 5m\_vs\_6m$\rightarrow$10m\_vs\_11m}.
\label{fig:generalization}
\end{figure*}
\begin{figure*}[ht]
    \centering
    \includegraphics[width=0.8\textwidth]{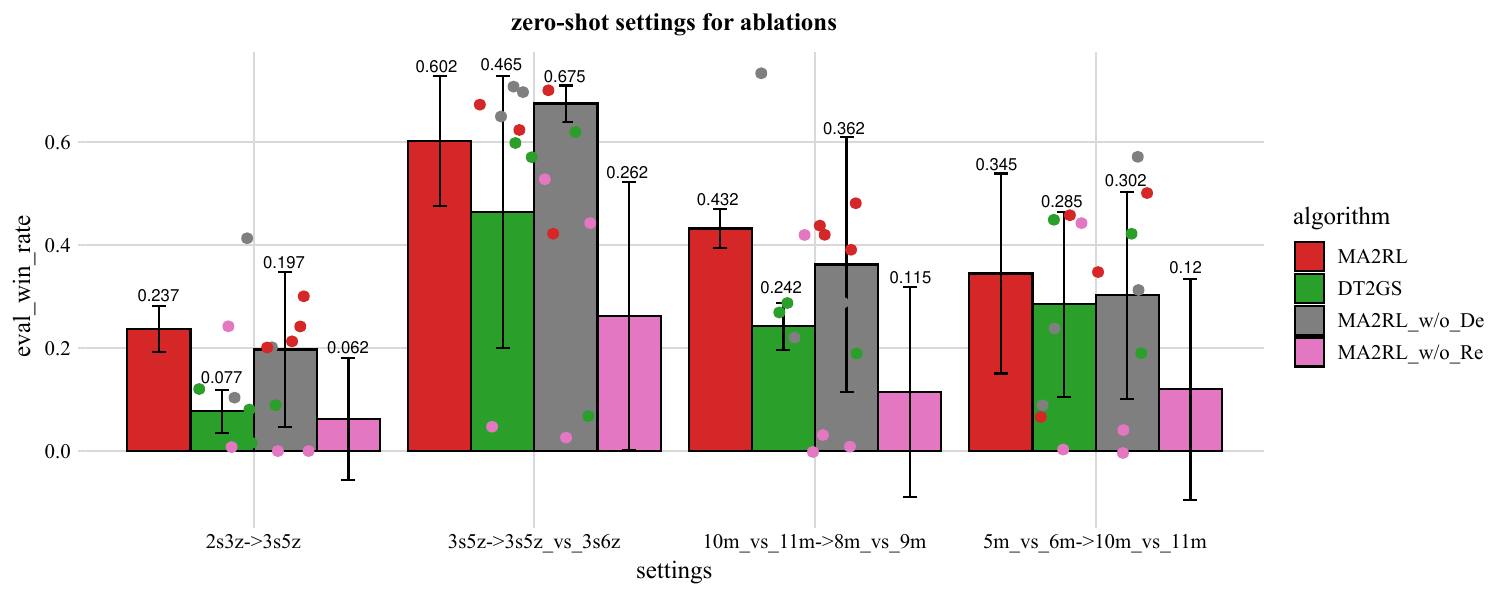}
    \caption{The zero-shot generalization capability of MA2RL and ablations.}
\label{fig:generalization_ablation}
\end{figure*}

\begin{figure*}[t]
    \centering
    \includegraphics[width=0.9\textwidth]{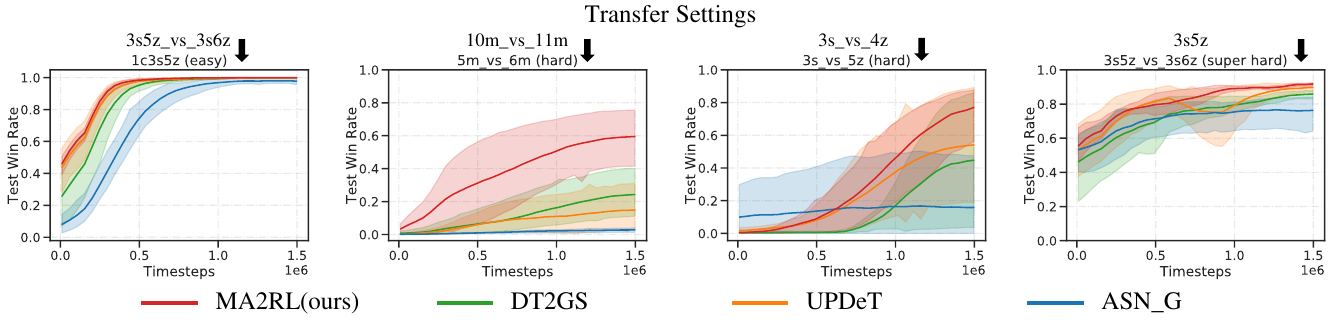}
    \caption{The transferability of MA2RL and its baselines, including DT2GS, UPDeT, ASN\_G, was compared across various source and target tasks.}
\label{fig:transfer}
\end{figure*}
\begin{figure*}[ht]
    \centering
    \includegraphics[width=0.9\textwidth]{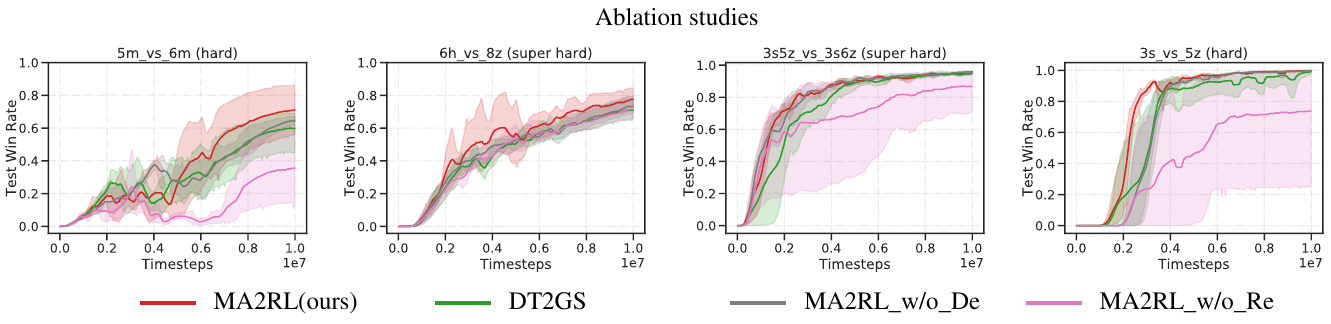}
    \caption{Ablation studies. MA2RL\_w/o\_De, it only removes the usage of latent representation of all masked entity-observation in attentive action decoder and MA2RL\_w/o\_Re, it is not reuse the VAE decoder in the attentive action decoder.}.
\label{fig:ablation}
\end{figure*}

\begin{table*}[ht]
    \centering
    \caption{The table shows a comparison of zero-shot generalization capability between baselines in MPE. "3->2" indicates that the source task is set by 3 agents and 3 landmarks, while the target task is 2 agents and 2 landmarks.}
    \begin{tabular}{|c|c|c|c|c|}
        \hline
        \textbf{source/target} & \textbf{MA2RL} & \textbf{DT2GS} & \textbf{UPDeT} & \textbf{ASN\_G} \\ 
        \hline
        3>2 & \textbf{-110.10 $\pm$ 4.61} & -113.06 $\pm$ 4.06 & -113.74 $\pm$ 3.35 & -157.55 $\pm$ 11.37 \\ 
        3>4 & \textbf{-273.96 $\pm$ 3.18} & -283.28 $\pm$ 7.35 & -288.11 $\pm$ 7.11 & -339.91 $\pm$ 8.75 \\ 
        3>5 & \textbf{-374.37 $\pm$ 5.83} & -377.99 $\pm$ 7.47 & -381.72 $\pm$ 8.34 & -485.12 $\pm$ 33.23 \\ 
        3>6 & \textbf{-477.23 $\pm$ 1.33} & -483.68 $\pm$ 8.53 & -488.97 $\pm$ 11.20 & -572.85 $\pm$ 11.64 \\ 
        \hline
    \end{tabular}
    \label{mpe_zeroshot1}
\end{table*}

\begin{table*}[t]
    \centering
    \caption{The figure shows a comparison of zero-shot generalization capability between baselines in MPE. "4->2" indicates that the source task is set by 4 agents and 4 landmarks, while the target task is 2 agents and 2 landmarks.}
    \begin{tabular}{|c|c|c|c|c|}
        \hline
        \textbf{source/target} & \textbf{MA2RL} & \textbf{DT2GS} & \textbf{UPDeT} & \textbf{ASN\_G} \\ 
        \hline
        4>2 & -111.27 $\pm$ 1.49 & \textbf{-111.07 $\pm$ 3.17} & -111.43 $\pm$ 4.25 & -117.15 $\pm$ 3.35 \\ 
        4>3 & \textbf{-185.78 $\pm$ 1.73} & -190.55 $\pm$ 4.69 & -192.18 $\pm$ 6.72 & -214.54 $\pm$ 9.71 \\ 
        4>5 & \textbf{-363.2 $\pm$ 2.76} & -368.46 $\pm$ 5.46 & -373.69 $\pm$ 8.73 & -431.35 $\pm$ 7.26 \\ 
        4>6 & \textbf{-468.67 $\pm$ 5.04} & -473.96 $\pm$ 7.85 & -479.84 $\pm$ 7.85 & -570.85 $\pm$ 18.52 \\ 
        \hline
    \end{tabular}
    \label{mpe_zeroshot2}
\end{table*}
Asymptotic performance and sample efficiency are the most fundamental evaluation metrics for a model. Therefore, we first evaluate MA2RL and the baselines on hard and superhard maps in SMAC. As shown in Fig.\ref{fig:single_task}, MA2RL obtains the best performance on most of the maps, particularly in hard and superhard maps such as 5m\_vs\_6m, 6h\_vs\_8z, 3s5z\_vs\_3s6z, 3s\_vs\_5z. It is worth noting that MA2RL outperforms all the baselines by the largest margin on the above-mentioned maps, which have numerous entities and suffer severely from partial observations. The results indicate that MA2RL can effectively model and infer information about masked entities through MAE. While DT2GS improves asymptotic performance and generalization capability through scalable semantics, the learning of semantics suffers severely from partial observations. MA2RL shows a faster increasing trend in win rate, which may be attributed to its inference regarding global states. Additionally, MA2RL exhibits lower variance in learning curves, indicating that modeling the masked entities can mitigate the impact of partial observability and enhance training stability.To further evaluate the proposed method in cooperative multi-agent settings, We conducted asymptotic performance experiments on Spread in MPE. As pepicted in Fig.~\ref{fig:asymptotic_mpe}, MA2RL outperforms all the baselines, whose curves rise earliest and fastest.

MA2RL decomposes observations into entity observations and applies MAE from the perspective of entities, naturally forming a cross-task generalizable model structure. Subsequently, we apply MA2RL in multi-task settings to further demonstrate the generalization and representation capability across tasks. We conduct two multi-task settings with different distributions of difficulty: \textit{stalker\_zealot}-series tasks (easy, easy, superhard) and \textit{marine}-series tasks (hard, hard, hard). In each multi-task setting, the policy interacts synchronously with multiple tasks and updates the policy using a mixed experience replay buffer. As shown in Fig.~\ref{fig:multi_task}, MA2RL exhibits the fastest convergence and highest win rate among the compared methods (DT2GS, UPDeT, ASN\_G). DT2GS and UPDeT closely follow MA2RL in asymptotic performance. ASN\_G fails to win in the hard multi-task setting. Similarly, in the single-task setting, MA2RL displays a lower variance across random seeds.

\subsection{Masked Autoencoders are generalizable Zero-shot \& Few shot learners}
\label{subsec:generalization}
Improving zero-shot generalization capability is the fundamental motivation behind our design. MA2RL employs MAE from the perspective of entities, aiming to alleviate the degradation of generalization caused by partial observability in MAS. We evaluate the generalization capability of MA2RL and the baselines in six different settings, each of which includes a target task that is either more difficult or equally difficult compared to the source task. Fig.~\ref{fig:generalization} shows that MA2RL outperforms all baselines in terms of zero-shot capability, especially in settings with uncertainty resulting from partial observability (e.g. 2s3z$\rightarrow$3s5z, 8m\_vs\_9m$\rightarrow$5m\_vs\_6m). These indicate that MA2RL's applying MAE from the entity perspective can effectively address the challenges of partial observation and varying observation/state/action spaces in MARL, thus effectively improving generalization. We also conducted zero-shot experiments on Spread in MPE and the results of the zero-shot experiments are presented in Table ~\ref{mpe_zeroshot1} and Table ~\ref{mpe_zeroshot2}. In the Table ~\ref{mpe_zeroshot1} and ~\ref{mpe_zeroshot2}, "4>2" in the column of "source/target" indicates that the source task is set by 4 agents and 4 landmarks, while the target task is set by 2 agents and 2 landmarks.

In conclusion, MA2RL is a generalizable zero-shot learner that can successfully transfer the ability to infer masked entity-observations to new tasks without requiring fine-tuning. Additionally, Fig.~\ref{fig:transfer} demonstrates that MA2RL has more transferabilty than baselines regarding asymptotic  performance and time to threshold.

\subsection{Ablations}
\label{subsec:ablation}
We carry out ablation studies to investigate how to leverage better the latent representation of all masked entity observations and access MAE's contributions in MARL. We compare MA2RL against four ablations:(1) \textit{MA2RL\_w/o\_De}, which excludes the utilization of the latent representation of all masked entity-observations in the attentive action decoder. (2) \textit{MA2RL\_w/o\_Re}, which does not reuse the VAE decoder in the attentive action decoder. (3) \textit{DT2GS}, which removes all components related to MAE. In these ablations, all other structural models are kept consistent strictly with the full MA2RL. Fig.~\ref{fig:ablation} shows ablation results on four representative maps. The results demonstrate that the design of reusing decoder in VAE and MAE are essential for MA2RL's capability. Specifically,  MA2RL\_w/o\_De slightly improves performance compared to DT2GS, confirming that MAE in MA2RL facilitates better assignment of individual skills to tackle complex tasks. Significantly, MA2RL\_w/o\_Re exhibits severe performance degradation compared to MA2RL, indicating that the reuse of the VAE decoder can effectively leverage the mapping relationship between latent space and observation space. The reuse of the VAE decoder effectively connects the two components.

To further verify the role of each module in MA2RL for policy generalization, we add zero-shot experiments in the ablation studies, as illustrated in Fig.~\ref{fig:generalization_ablation}. The results demonstrate the effectiveness of different components in MA2RL from both the aspects of generalization and asymptotic performance:
\begin{itemize}
    \item generalization: MA2RL significantly outperforms DT2GS by a large marin, demonstrating the excellent generalization of MA2RL. The comparison between MA2RL and MA2RL\_w/o\_De shows that using the inferred mask information can learn skill semantics with stronger generalization.
    \item asymptotic performance: MA2RL does not introduce additional hyperparameters. All parameter settings are consistent with the baselines, thus the designed MAE can improve the asymptotic performance to some extent. Additionally, MA2RL can greatly enhance generalization without damaging the asymptotic performance, and even improve the asymptotic performance in some difficult tasks (e.g. 5m\_vs\_6m, 6h\_vs\_8z).
\end{itemize}

\section{Conclusion and Future work}
\label{Conclusions}
To tackle the challenges of generalization in MARL caused by partial observation, this paper proposes a generalizable framework that makes the first attempt to extend MAE in the context of generalization in MARL. Experimental results and ablation studies verify the extraordinary performance of MA2RL. However, our framework does not consider offline settings more relevant to real-world scenarios, as online interaction can be costly. And our framework is limited to various scenarios within one type of game. In the future, extending MA2RL to create a single generalist agent capable of generalizing across different types of games is a promising direction we hope to explore.

\bibliographystyle{IEEEtran}
\bibliography{references}
\end{document}